\documentclass[authoryear,preprint,review]{elsarticle}
\usepackage{amssymb}
\usepackage{hyperref}

\newcommand{\cmt}[1]{}
\newcommand{\Rs}{R_{sub}}
\newcommand{\Rt}{R_{tot}}
\newcommand{\fo}{f_{1}(p)}
\newcommand{\de}{\partial}
\newcommand{\pj}{p_{inj}}
\newcommand{\px}{p_{max}}

\newcommand{\di}{{\rm d}}
\newcommand{\degK}{\,\rm{ K}}
\newcommand{\simgt}%
       {\,\hbox{\lower0.6ex\hbox{$\sim$}\llap{\raise0.6ex\hbox{$>$}}}\,}

   \newcommand{\simlt}%
       {\,\hbox{\lower0.6ex\hbox{$\sim$}\llap{\raise0.6ex\hbox{$<$}}}\,}
       
\begin{document}
\begin{frontmatter}

\title{Non-linear diffusive shock acceleration with free escape boundary: analytical solution}

\author{D. Caprioli}
\ead{caprioli@arcetri.astro.it}

\author{E. Amato}
\ead{amato@arcetri.astro.it}

\author{P. Blasi}
\ead{blasi@arcetri.astro.it}

\address{INAF/Osservatorio Astrofisico di Arcetri, Largo E. Fermi, 5 - 50125 Firenze, Italy}
\address{Kavli Institute for Theoretical Physics, Kohn Hall, Santa Barbara (CA) 93106, US}

\begin{abstract}
We present here a semi-analytical solution of the problem of particle acceleration at non-linear shock waves with a free escape boundary at some location upstream. 
This solution, besides allowing us to determine the spectrum of particles accelerated at the shock front, including the shape of the cutoff at some maximum momentum, also allows us to determine the spectrum of particles escaping the system from upstream. 
This latter aspect of the problem is crucial for establishing a connection between the accelerated particles in astrophysical sources, such as supernova remnants, and the cosmic rays observed at the Earth.
An excellent approximate solution, which leads to a computationally fast calculation of the structure of shocks with an arbitrary level of cosmic ray modification, is also obtained.
\end{abstract}


\end{frontmatter}

\section{Introduction}

The importance of the process of particle acceleration in astrophysical shock waves for the origin of cosmic rays (CRs) is now generally acknowledged but several weak points remain in the theory when one tries to establish a connection between accelerated particles and cosmic rays observed at Earth. The main problem is related to the difficulties at assessing the role of escaping particles: while supernova remnants (SNRs) are often invoked as the main sources of Galactic CRs, at least up to the knee, their ability to generate CRs with the spectrum observed at Earth is all but proven. If particles are trapped in the expanding shell during the Sedov-Taylor phase, adiabatic energy losses prevent the release in the interstellar medium of particles with energies in the knee region. 
If SNRs are the main sources of CRs up to the knee, ongoing escape of particles from the upstream region is required during the Sedov-Taylor phase. The spectrum of these escaping particles is expected to be peaked around the maximum momentum that can be reached in the accelerator at a given time. The actual spectrum of CRs from an individual SNR is therefore the convolution over time of these peaked spectra. Despite the crucial importance of escaping particles, their role in cosmic ray modified shocks has received scarce attention so far, with some noticeable exceptions (see for instance the work by \cite{bere04,pz05,lee08,reville,escape}). One of the difficulties from the technical point of view is that it is not clear which particles do actually escape the system. While from the mathematical point of view, escape can be modeled by requiring the existence of a free escape boundary upstream, from the physical point of view the issue remains that the position of this boundary is related to poorly understood details of the problem, especially the ability of particles to self-generate their own scattering centers. The position of the free escape boundary could coincide with a location upstream of the shock where particles are no longer able to scatter effectively and return to the shock. This however would lead to an anisotropic distribution function of the accelerated particles, that can no longer be described by the standard diffusion-convection equation. Moreover, while waves can be generated both resonantly \cite[]{skillinga} and non-resonantly \cite[]{bell2004}, particles can scatter effectively only with resonant waves. This adds to the complexity of the problem, in that one might have amplified magnetic fields of large strength but on scales which do not imply effective scattering of the highest energy particles. 

In the absence of a better description of this phenomenon, so far the best way to mimic the escape is to impose a {\it reasonable} location for a free escape boundary and calculate the escape flux as derived from the diffusion approximation. Here we present an exact semi-analytical solution of this problem for shocks with arbitrary cosmic ray induced modification. We also propose a simple approximate expression which turns out to be an excellent approximation to the exact solution. The approach presented here allows us to calculate the spectrum of accelerated particles at any location upstream and downstream of the shock and the spectrum of escaping particles, in the assumption that a quasi-stationary situation is reached at any given time. Clearly within this approach the maximum momentum achieved by the particles is not imposed by hands but rather obtained self-consistently from the condition of free escape at $x_{0}$. 

This is not the first attempt in the literature at investigating the problem of free escape from a shock region: the problem was recently faced numerically by \cite{reville} and \cite{za09}, who specialized their calculation to the case of SNR RX J1713.7-3946. \cite{kj95,kj06} investigated the problem of particle acceleration at a modified shock with free escape through a time-dependent finite differences scheme with adaptive mesh refinement of the grid. 
A Monte Carlo technique was adopted by e.g. \cite{je91,elli06} to have a handle on the escape flux of particles from the shock. 
It is worth stressing however that these numerical methods require computation times for a given set of parameters which range between several hours, for the Monte Carlo technique, and several days for the time dependent calculations of \cite{kj95,kj06}. These times should be compared with a typical computation time of 1-2 minutes (on a laptop) required for the semi-analytical method discussed here or previous versions of it, in which a boundary condition in momentum was adopted \cite[]{blasi1,blasi2,amato1,amato2}. The issue of computation time becomes crucial when these calculations are embedded in hydrodynamical codes for the evolution of SNRs. 

The paper is organized as follows: in \S \ref{sec:exact} we obtain the implicit exact solution of the problem with a given free escape boundary condition; the approximate solution, along with its comparison with the exact one, is presented in \ref{sec:approx}. We conclude in \S \ref{sec:concl}.

\section{Exact solution}
\label{sec:exact}

We start from the stationary, non-relativistic, one dimensional diffusion-convection equation for the isotropic part of the distribution function of accelerated particles $f(x,p)$ \citep[see e.g.\ ][]{skillinga}:
\begin{equation}
u(x)\frac{\de f(x,p)}{\de x}=\frac{\de}{\de x}\left[D(x,p)\frac{\de f(x,p)}{\de x}\right]+\frac{p}{3}\frac{\di u(x)}{\di x}\frac{\de f(x,p)}{\de p}+Q(x,p)\,,
\label{eq:trans}
\end{equation}
where $D(x,p)$ is the diffusion coefficient, with arbitrary dependence on both position and momentum, $Q(x,p)$ is the injection rate and $u(x)$ is the fluid velocity in the shock frame. Here, for the sake of clarity, we neglect the velocity of the scattering centres, which is typically of order of the Alfv\'en velocity $v_{A}$, with respect to the fluid. The generalization to the case of small Alfv\'enic Mach number may be easily obtained following the procedure discussed in \citep[sec.~3]{lungo}. We solve this equation with the upstream boundary condition $f(x_{0},p)=0$, which mimics the presence of a free-escape boundary placed at a distance $x_{0}$ upstream of the shock (placed at $x=0$). The downstream region corresponds to $x>0$. Hereafter, we will label with the subscript $0,~1,~2$ quantities calculated at $x_{0}$, $x=0^{-}$ and $x=0^{+}$ respectively.

An implicit solution of Eq.~\ref{eq:trans} in the upstream region may be found by generalizing the approach used by \cite{malkov1}. Eq.~\ref{eq:trans} can be spatially integrated in the upstream region from $x$ to $x_{0}$ leading to:
\begin{equation}
D(x,p)\frac{\partial f}{\partial x}-u(x) f(x,p)=\left[D(x,p)\frac{\partial f}{\partial x}\right]_{x_0}-\frac{1}{3p^2}\int_{x_0}^x dx'\ \frac{du}{dx'}\frac{\partial}{\partial p}\left[p^3 f(x',p)\right]\, .
\label{eq:interm}
\end{equation}
The solution of the homogneneous equation associated to Eq.~\ref{eq:interm} still reads $f_h(x,p)=\exp{[\psi(x,p)]}$ with
\begin{equation}
\psi(x,p)=-\int_{x}^{0}\di x'\frac{u(x')}{D(x',p)}\ ,
\end{equation}
from which the general solution follows as:
\begin{equation}
f(x,p)=\fo \exp\left[\psi(x,p)\right]
	\left\{1+u_{0}\int_{x}^{0} \di x'\frac{\exp\left[-\psi(x',p)\right]}{D(x',p)}
	\left[\frac{\phi_{esc}(p)}{u_{0}\fo}+Z(x',p)\right]\right\}\,,
\end{equation}
where
\begin{equation}\label{eq:Zxp}
Z(x,p)=\frac{1}{u_{0}\fo}\int_{x_{0}}^{x}\di x' \frac{\di u(x')}{\di x'}\left[f(x',p)+\frac{p}{3}\frac{\de f(x',p)}{\de p}\right]\,,
\end{equation}
$\fo=f(0,p)$ is the particle spectrum at the shock location, and 
\begin{equation}
	\phi_{esc}(p)=-\left[D(x,p)\frac{\partial f}{\partial x}\right]_{x_0}
\end{equation}
is the flux of particles escaping from the shock across the surface at $x=x_0$ (escape flux).
This solution does not explicitly show that $f(x_{0},p)=0$, although this condition has been clearly used in passing from Eq.~\ref{eq:trans} to Eq.~\ref{eq:interm}. On the other hand, the condition $f(x_{0})=0$ directly leads to an interesting expression for the escape flux $\phi_{esc}(p)$, as soon as the transport equation is integrated between $x_0$ and the shock location:
\begin{equation}
\phi_{esc}(p)=-\fo \left\{ 1+ u_{0}\int_{x_{0}}^{0} \di x\frac{\exp\left[-\psi(x,p)\right]}{D(x,p)}Z(x,p)\right\} \left\{\int_{x_{0}}^{0} \di x \frac{\exp\left[-\psi(x,p)\right]}{D(x,p)}\right\}^{-1}\,.
\end{equation}
It is finally convenient to introduce the dimensionless functions $K(x,p)$ and $W(x,p)$ defined respectively as
\begin{equation}\label{eq:Kxp}
K(x,p)=u_{0}\int_{x}^{0} \di x'\frac{\exp\left[-\psi(x',p)\right]}{D(x',p)} Z(x',p)
\end{equation}
and
\begin{equation}
W(x,p)=u_{0}\int_{x}^{0} \di x' \frac{\exp\left[-\psi(x',p)\right]}{D(x',p)},
\end{equation}
so that the solution of the transport equation becomes:
\begin{equation}\label{eq:solx}
f(x,p)=\fo \exp\left[\psi(x,p)\right]\left\{1+K(x,p)
		-\frac{W(x,p)}{W_{0}(p)}\left[1+K_{0}(p)\right]\right\}
\end{equation}
and the escape flux:
\begin{equation}\label{eq:solfesc}
\phi_{esc}(p)=-u_{0}\fo \frac{1+K_{0}(p)}{W_{0}(p)}\,.
\end{equation}

At this point we follow the procedure described e.g.\ in \cite{blasi1,blasi2} of integrating Eq. \ref{eq:trans} across the shock and between $x_{0}$ and $0^{-}$ in order to derive an equation for $\fo$, i.e.\ :
\begin{equation}\label{eq:shock}
\frac{p}{3}\frac{\de f_{1}}{\de p}(u_{2}-u_{p})=\fo\left[ u_{p} +\frac{p}{3}\frac{\di u_{p}}{\di p}\right]+ \phi_{esc}(p)-Q_{1}(p)\,,
\end{equation}
where we introduced the mean velocity effectively felt by a particle with momentum $p$ in the upstream region:
\begin{equation}\label{eq:Up}
u_{p}(p)=u_{1}-\frac{1}{\fo}\int_{x_{0}}^{0}\di x \frac{\di u(x)}{\di x}f(x,p).
\end{equation}
Following \citep[see][]{bgv05} we write the injection term as
\begin{equation}
Q(x,p)=Q_{1}(p)\delta(x)=\frac{\eta n_{0}u_{0}}{4\pi \pj^{2}}\delta(p-\pj)\delta(x)\,,
\end{equation}
where $\eta$ is the fraction of particles crossing the shock and injected in the acceleration process, and $\pj$ is the injection momentum. As discussed by \cite{bgv05}, we write the injection momentum as a multiple of the thermal momentum of particles downstream, $\pj=\xi_{inj}p_{th,2}$. In the assumption that the thermal particles downstream have a Maxwellian spectrum, the fraction $\eta$ is uniquely determined by the choice of $\xi_{inj}$.

We also introduce the normalized fluid velocity $U=u/u_{0}$ and the normalized escape flux $\Phi_{esc}=\phi_{esc}/(u_{0}f_{1})$, as well as the compression ratios at the subshock $\Rs=u_{1}/u_{2}$ and between $x_{0}$ and downstream $\Rt=u_{0}/u_{2}$. The solution of Eq.~\ref{eq:shock} then reads:
\begin{equation}\label{eq:solshock}
\fo=\frac{\eta n_{0}}{4\pi \pj^{3}}\frac{3\Rt}{\Rt U_{p}(p)-1}
	\exp\left\{-\int_{\pj}^{p}\frac{\di p'}{p'}\frac{3\Rt\left[U_{p}(p')-\Phi_{esc}(p')\right]}{\Rt U_{p}(p')-1}\right\}.
\end{equation}

It is easy to check that in the test-particle limit $K(x,p)=0,\ U_{p}(p)= 1$ and the standard solution \citep[see e.g.\ ][]{escape} is recovered in eqs.~\ref{eq:solx}, \ref{eq:solfesc} and \ref{eq:solshock}.

\section{An approximate solution}
\label{sec:approx}

In this section we use a heuristic argument to derive an approximate solution of the problem. An {\it a posteriori} comparison with the exact solution derived above shows an excellent agreement in all cases considered.

Let us consider the function $K(x,p)$ defined in Eq.~\ref{eq:Kxp}.  At any given momentum $p$, the distribution function can be regarded as approximately constant ($f(x,p)\simeq \fo$) for $x<x_{p}\simeq D(p)/u(x_{p})$ and exponentially suppressed for $x>x_{p}$. Hence, for $x\ll x_{p}$ we have $Z(x,p)\simeq U(x_{p})-U(x)<1$ but also $\psi(x,p)\ll 1$. This leads to $K(x,p)\approx (u_{0}x/D(p)) Z(x,p)\ll 1$. On the other hand, for $x\gg x_{p}$, $Z(x,p)\to 0$ so that both $\exp\left[\psi(x,p)\right]K(x,p)$ and $K_{0}(p)/W_{0}(p)$ tend to 0. For $x\sim x_{p}$, the situation is less clear, but one can expect that $Z(x,p)\ll 1$ because $x\to x_{p}$ and $f_{1}(p)$ starts feeling the exponential suppression. This suggests that we can neglect $K(x,p)$ with respect to unity in Eq.~\ref{eq:solx}, although clearly this conclusion needs to be checked {\it a posteriori} against the exact solution.

The following recipe is thus proposed as an approximation of the exact solution of the transport equation:
\begin{equation}\label{eq:app}
f(x,p)=\fo\exp\left[-\int_{x}^{0}\di x'\frac{u(x')}{D(x',p)}\right]	\left[ 1-\frac{W(x,p)}{W_{0}(p)}\right];
\end{equation}
\begin{equation}
\phi_{esc}(p)=- \frac{u_{0}\fo}{W_{0}(p)}\,.
\end{equation}
This expression tends to the correct test-particle limit, as one can easily verify. A point that is worth highlighting is that the distribution function at the shock, $\fo$, is sensible to the assumed spatial dependence only through the function $U_{p}$, and is therefore weakly affected by whether the approximate or the exact solution is adopted. 

We want to stress an important point: in the case of boundary condition in momentum, namely when the maximum momentum is fixed, the procedure above would lead us to the functional form $f(x,p)=\fo\exp\left[-\int_{x}^{0}\di x'\frac{u(x')}{D(x',p)}\right]$, which is slightly different (and simpler) than any ansatze previously proposed in the literature \citep[see for instance][]{malkov1,bac07}. 
It turns out that the approximation found here, which is the simplest possible extrapolation of test-particle theory, gives a solution that is basically undistinguishable from the exact one. In other words, both in the case of boundary condition in momentum (fixed $p_{max}$) or free escape boundary (fixed $x_{0}$), the best description of the spatial distribution of accelerated particles is provided by the simplest possibility which automatically satisfies all the relevant limits and the boundary condition at the shock. 

A full solution of the system of conservation equations for mass, momentum and energy, coupled with the diffusion-convection one, is obtained following the iterative procedure described by \cite{amato1} and by \cite{lungo}, when the generation of magnetic turbulence via streaming instability \citep[see e.g.][]{skillinga} and its dynamical feedback \citep{jumpl} are taken into account. The only difference here is that there is no need to fix a maximum momentum by hand, since the distribution function gets intrinsically suppressed above a certain $\px$ as a consequence of the escape at $x=x_{0}$.

The iterative method can be summarized as follows. Let us consider the momentum conservation equation, normalized to $\rho_{0}u_{0}^{2}$:
\begin{equation}\label{eq:momentum}
U(x)+P_{c}(x)+P_{w}(x)+P_{g}(x)=1+\frac{1}{\gamma M_0^2}\,,
\end{equation}
where we introduced the normalized pressure in cosmic rays:
\begin{equation}\label{eq:pc}
	P_{c}(x)=\frac{4\pi}{3\rho_0u_0^2}\int_{\pj}^{\infty}dp~ p^3~ v(p) ~f(x,p)\,,
\end{equation}
the normalized pressure in magnetic turbulence generated via resonant streaming instability
\citep[see][Eq.~42]{lungo}:
\begin{equation}\label{eq:pw}
	P_{w}(x)=\frac{v_A}{4u_0}\frac{1-U(x)^2}{U(x)^{3/2}},
\end{equation}
and the normalized pressure of the background gas with adiabatic index $\gamma$:
\begin{equation}\label{eq:pg}
P_{g}(x)=\frac{U(x)^{-\gamma}}{\gamma M_{0}^{2}}\,.
\end{equation}
The last expression holds provided the heating in the precursor is purely adiabatic: its generalization to cases with some turbulent heating is however straightforward \citep[see e.g.][sec. 6]{lungo}.

We start from a guess value for $U_{1}=\Rs/\Rt$, which uniquely determines $P_{w1}$, $P_{g1}$ and $P_{c1}$ via Eqs.~\ref{eq:pw}, \ref{eq:pg} and \ref{eq:momentum}. 
We notice that, once $U_{1}$ is fixed, $\Rs$ and $\Rt$ can be worked out separately from the conservation equations in the precursor obtaining \citep[see][eq.~16]{lungo}
\begin{equation}\label{rsrt}
\Rt^{\gamma+1}=\frac{M_0^2\Rs^\gamma}{2}\frac{\gamma+1-\Rs(\gamma-1)}  
{1+P_{w1}/P_{g1}\left[1+\Rs\left(2/\gamma-1\right)\right]}
\end{equation}
and in turn
\begin{equation}
\Rt=\frac{\lambda-\sqrt{\lambda^{2}-8P_{w1}U_{1}(\gamma+1)(\gamma-2)}}{4U_{1}P_{w1}(\gamma-2)};
\quad \Rs=U_{1}\Rt
\end{equation}
where $\lambda=2\gamma(P_{g1}+P_{w1})+U_{1}(\gamma-1)$.

At this point, we start with a test-particle guess for $f(x,p)$, normalized in order to account for the obtained $P_{c1}$, and calculate $P_{c}(x)$ by using Eq.~\ref{eq:pc} and then $U(x)$ through Eq.~\ref{eq:momentum}. This updated velocity leads to a new $P_{w}(x)$ and hence to $\delta B(x)=\sqrt{8\pi\rho_{0}u_{0}^{2}P_{w}(x)}$ which is used to update $D(x,p)$.

Now we can calculate a new $f(x,p)$ as a function of the old distribution function and of the new $U(x)$ and $D(x,p)$, according to Eq.~\ref{eq:solshock} and Eq.~\ref{eq:solx} (or Eq.~\ref{eq:app}). The new $f(x,p)$ is again normalized to $P_{c1}$ and the procedure above is iterated until convergence is reached, i.e.\ until $f(x,p)$ and its normalization factor do not change between two successive steps.

For an arbitrary value of $U_{1}$, however, the required normalization factor will be different from 1, thus the process is restarted with a different choice of $U_{1}$ until no further normalization is needed. The distribution function calculated with the value of $\Rs/\Rt$ obtained in this way is by construction the solution of both transport and conservation equations.

We consider here two cases: a test-particle-like one (inefficient acceleration) and a strongly modified one. In the first case, we choose $\xi_{inj}=4.3$, corresponding to a fraction of injected particles $\eta\simeq 1.7\times 10^{-6}$, while in the second case we choose $\xi_{inj}=3.3$, corresponding to $\eta\simeq 1.2\times 10^{-3}$. Moreover, in the inefficient case we assume Bohm diffusion in the background magnetic field $B_{0}=5\mu G$, while in the strongly modified case, having in mind the case of shocks in SNRs, we adopt a Bohm-like diffusion coefficient calculated in the magnetic field which is self-generated through resonant streaming instability by accelerated particles \cite[]{amato2}, namely $D(x,p)=\frac{c}{3}\frac{pc}{e\delta B(x)}$. The other parameters are chosen as follows: the shock velocity is $u_{0}=$5000 km/s, the free escape boundary is located at $x_{0}=0.15$ pc, and finally the background density (temperature) is $\rho_{0}=0.1 m_{p}$cm$^{-2}$ ($T_{0}=10^{5}\degK$), corresponding to a sonic Mach number $M_{0}\simeq 135$. Again, these choices are inspired by the values expected in SNRs.

\begin{figure}
   \centering
   \includegraphics[width=1\textwidth]{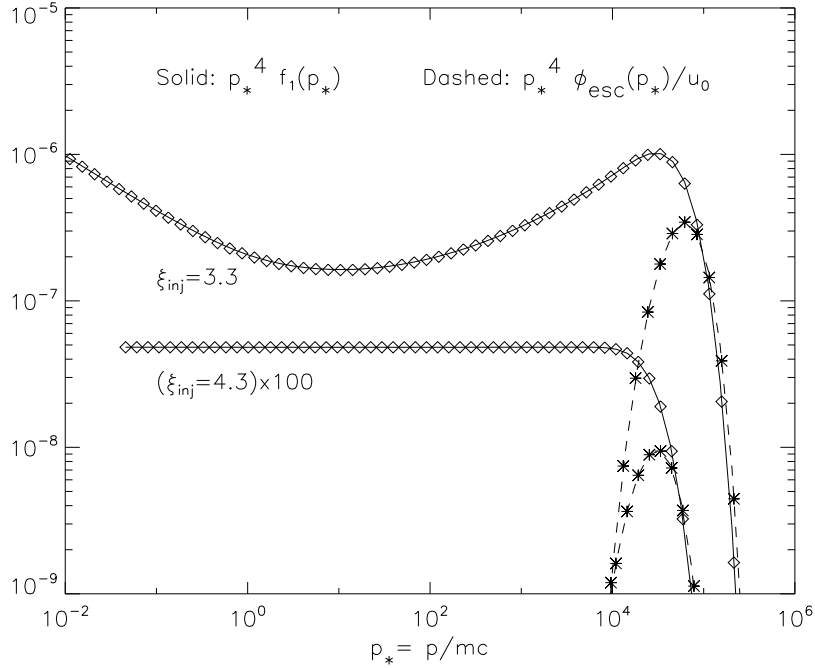} 
   \caption{Particle spectra at the shock and escape flux in the test-particle-like case multiplied by a factor 100 ($\xi=4.3$) and in a strongly modified case ($\xi=3.3$).
   Symbols correspond to the approximate solution given by Eq.~\ref{eq:app}, while lines correspond to the exact solution (Eq.~\ref{eq:solx}).}
   \label{fig:spectrum}
\end{figure}

\begin{figure}
   \centering
\includegraphics[width=1\textwidth]{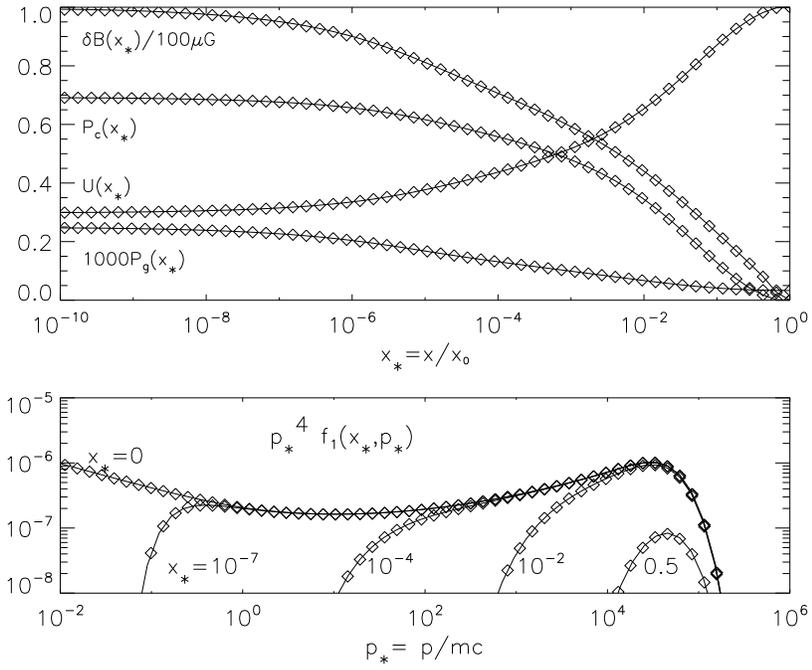}
\caption{Upstream hydrodynamical quantities (top) and cosmic ray distribution function at $x_{*}=x/x_0=10^{-7},10^{-4},10^{-2},0.5$ (bottom), in the $\xi=3.3$ case. 
   In any panel symbols correspond to the approximate solution and lines correspond to the exact solution.}
   \label{fig:hydro}
\end{figure}

In Fig.~\ref{fig:spectrum} we plot the particle spectrum at the shock and the escape flux in a test-particle-like case (multiplied by a factor 100, lower curves) and in a strongly modified one (upper curves). 
In Fig.~\ref{fig:hydro}, instead, we show the hydrodynamical quantities in the upstream region (top panel), and the distribution function at some given upstream positions, namely at $x/x_0=10^{-7},10^{-4},10^{-2},0.5$ (bottom panel), all referred to the modified case.
All the curves in the Figs.~\ref{fig:spectrum} and \ref{fig:hydro} refer to the exact solution. The approximate solution given by Eq.~\ref{eq:app} leads to the results shown with symbols in both figures. 
One can easily realize that the agreement between the results obtained with the exact and the approximate solution is excellent, beyond any expectation. 
Moreover, the case of inefficient acceleration reduces exactly to the test-particle case, as shown by the lower curves in Fig.~\ref{fig:spectrum}.

The efficient case shows the typical features of cosmic ray modified shocks, with a concavity in the spectrum induced by the precursor in the upstream fluid \citep[see e.g.][for a comprehensive review]{maldrury}. 
In the present case the total compression coefficient is $\Rt\simeq 10.6$ as a result of the pressure in cosmic rays (about $66$ per cent of the bulk pressure at the shock) and of the dynamical backreaction of the amplified magnetic field, since upstream the magnetic pressure dominates over the gas pressure ($P_{w1}/P_{g1}\simeq45$), as described in \cite{jumpl}. In this case the energy carried away by escaping particles represents about $37$ per cent of the bulk energy flux.

It is interesting to notice that in the case of efficient acceleration, despite the fact that the magnetic field amplification induced by accelerated particles at the shock is of order $\delta B/B_{0}\sim 20$, the maximum momentum which is implied by the free escape boundary condition at $x=x_{0}$ is only a factor $\sim 2$ higher than in the inefficient case (we recall that in this latter case the diffusion coefficient is Bohm-like in the background magnetic field $B_{0}$). 
This apparently counter intuitive result is in fact simple to understand: due to the dynamical reaction of the accelerated particles, the effective fluid velocity felt by particles in the precursor is $U_{1}\simeq0.3$, which implies a slower acceleration rate and lower maximum momentum; moreover the fact that in the efficient scenario the magnetic field is self-generated implies that most of it is concentrated around the shock, while the (turbulent) magnetic field responsible for particle diffusion close to $x_{0}$ is in fact much smaller than $B_{0}$ ($\delta B$ turns out to be smaller than $B_{0}$ for $x\simlt 0.5x_{0}$). As a consequence of these two facts, the maximum momentum does increase in the modified case with respect to the inefficient one, but less than the naive expectation would suggest: in fact, as far as our investigation of the parameter space has gone so far, the maximum momentum does scale linearly with the position of the free escape boundary $x_{0}$ but not with the strength of the amplified magnetic field.

\section{Discussion and Conclusions}
\label{sec:concl}

Here we discussed the first semi-analytical exact solution of the problem of particle acceleration in non-relativistic shocks with a free escape boundary, when non-linear effects induced by the dynamical reaction of accelerated particles and by the amplification and dynamical feedback of the magnetic field are taken into account. In addition to the exact solution, which is rather cumbersome to implement in numerical calculations, we also proposed a simple but excellent approximation to the exact solution. This approximate solution catches all the main Physics ingredients of the problem and is computationally very convenient. We checked this approximate solution versus the exact solution and the agreement, both in terms of the spectrum of accelerated particles at the shock and in terms of the spatial distribution of accelerated particles in the precursor, is excellent. As a consequence, also the shock structure in terms of spatial dependence of the hydrodynamical quantities and of the self-generated magnetic field is perfectly reproduced. The escape fluxes and spectra are also in stunning agreement. 

The ability at providing not only the spectrum of accelerated particles, but also the spectrum of particles escaping through the free escape boundary located at a position $x_{0}$ upstream, is exactly what makes the solutions presented here (both the exact one and the approximate one) especially valuable. In a realistic situation, such as the expanding shock front associated with a supernova remnant, the existence of a free escape boundary leads to a maximum momentum of the accelerated particles which depends on time and in general decreases with time during the Sedov-Taylor phase, if the magnetic field is generated by the accelerated particles through streaming instability \cite[]{escape}. It follows that the convolution in time of the instantaneous escape flux leads to the formation of a complex spectrum which is no longer peaked around a specific momentum. 

From the physical point of view the main uncertainty related to this type of calculation is 1) in the nature of the self-generated waves and their interaction with accelerated particles and 2) in the determination of the location of the free escape boundary based on first principles. These two issues, clearly related to each other, are not easily solvable at the present time and a phenomenological approach is the only one we can afford to adopt. 

From the mathematical point of view, the solution presented here is an important step forward in the description of the process of particle acceleration in astrophysical shocks, especially in SNRs. The limitation that remains is that the solution assumes that the system is able to reach a quasi-stationary configuration at any given time. Despite this limitation these methods are of the greatest importance in order to have an appropriate description of the acceleration process in complex astrophysical objects such as SNRs. Other methods, all numerical in nature, have in fact typical running times that range between several hours and several days for a given set of parameters, compared with $\cal O$(minutes) required by the semi-analytical approach presented here.

\section*{Acknowledgments}
This work was partially supported by MIUR (under grant PRIN-2006) and by ASI through contract ASI-INAF I/088/06/0. This research was also supported in part by the National Science Foundation under Grant No. PHY05-51164.
\label{lastpage}

\end{document}